# Hierarchical clustering of mixed-type data based on barycentric coding

Odysseas Moschidis[1], Angelos Markos[2], Theodore Chadjipadelis[3]

[1]Department of Business Administration, University of Macedonia, Greece

[2]Department of Primary Education, Democritus University of Thrace, Greece

[3]Department of Political Sciences, Aristotle University of Thessaloniki, Greece

Corresponding author: Angelos Markos, School of Education, Department of Primary Education, Democritus University of Thrace, N. Hili, GR 68131, Alexandroupolis, Greece.

Email: amarkos@eled.duth.gr

## Statements and Declarations

**Competing interests**

On behalf of all authors, the corresponding author states that there is no conflict of interest.

**Funding**

The authors declare that no funds, grants, or other support were received during the preparation of this manuscript.

## Abstract

Clustering of mixed-type datasets can be a particularly challenging task as it requires taking into account the associations between variables with different level of measurement, i.e., nominal, ordinal and/or interval. In some cases, hierarchical clustering is considered a suitable approach, as it makes few assumptions about the data and its solution can be easily visualized. Since most hierarchical clustering approaches assume variables are measured on the same scale, a simple strategy for clustering mixed-type data is to homogenize the variables before clustering. This would mean either recoding the continuous variables as categorical ones or vice versa. However, typical discretization of continuous variables implies loss of information. In this work, an agglomerative hierarchical clustering approach for mixed-type data is proposed, which relies on a barycentric coding of continuous variables. The proposed approach minimizes information loss and is compatible with the framework of correspondence analysis. The utility of the method is demonstrated on real and simulated data.

**Keywords** Hierarchical cluster analysis · Mixed-type data · Ward clustering · Chi-square distance

## 1 Introduction

Many real-world datasets contain both continuous and categorical variables. Cluster analysis is often applied to such mixed-type data to group similar observations for further data analysis. Some popular application areas include business and marketing (Morlini and Zani 2010), health and biology (McParland et al. 2017) and social stratification (Hennig and Liao 2013). A recent review of clustering methods for mixed-type data can be found in Ahmad and Khan (2019). Clustering methods can be generally classified into model-based and non-model-based. Model-based clustering methods, also known as finite-mixture models, utilize probability models, whereas non-model-based approaches, such as hierarchical and partitioning methods are distance-based or algorithmic approaches, without assuming any underlying probability model. For a recent review that focuses on non model-based clustering approaches for mixed-type data, see van de Velden, Iodice D'Enza and Markos (2019). In this work, our focus is on hierarchical clustering of mixed-type data.

Hierarchical clustering methods rely on calculating a (dis)similarity matrix, based on the (dis)similarity between each pair of observations on the variables. A linkage criterion is defined to determine the distance between sets of observations as a function of the pairwise distances between observations. The hierarchy of clusters is created in a top to down (or vice versa) order. The choices of the (dis)similarity metric and the linkage criterion largely influence the final clustering solution. Most distance metrics, and hence the hierarchical clustering methods, work either with continuous-only or categorical-only data. In applications, however, observations are often described by a combination of both continuous and categorical variables. We can distinguish two general strategies for hierarchical clustering of mixed-type data: (a) construct a (dis)similarity measure that can be applied directly to mixed-type data, and (b) recode the continuous variables as categorical ones (or vice-versa) and then apply a suitable hierarchical clustering method. We discuss both strategies below.

A popular strategy for hierarchical clustering of mixed-type data is to construct a (dis)similarity measure that can be applied directly to mixed data (see, e.g., Hennig and Liao 2013). In this case, variables are divided into different subsets, according to their measurement level, i.e., one with nominal, one with ordinal and one with interval-scaled variables. An appropriate (dis)similarity measure is calculated for each subset and the results are combined into a single measure. Gower's similarity coefficient is one of the most popular measures of proximity for mixed data types (Gower 1971). Gower's coefficient combines the simple matching coefficient for categorical variables with the range-normalized absolute difference for continuous variables. The similarities of the categorical variables and the continuous variables are combined to compute the overall similarity between two observations; a hierarchical agglomerative clustering on the distance matrix can be used to create clusters. The Similarity Based Agglomerative Clustering (SBAC) algorithm (Li and Biswas 2002) is based on Goodall's similarity measure, that gives a greater weight to rare variable categories in similarity computations. An agglomerative algorithm is used to derive a dendrogram, and then using a heuristic technique a partition of the data is extracted. Hsu et al. (2007) proposed a hierarchical clustering algorithm for mixed-type data, where the similarity measure is based on the concept of distance hierarchy. For categorical variables, this hierarchy is a tree where each node represents a possible value of the variable and every link is valued by a weight representing the distance between these values. However, to create distance hierarchies, domain knowledge is required. For continuous variables, the hierarchy is represented by only two nodes corresponding to the minimum and maximum of the variable. The overall distance between two observations is calculated as the sum of their associated distance hierarchies. Ahmad and Dey (2007) proposed a K-means variant for mixed-type data, combining the squared Euclidean distance for continuous variables with a special distance for categorical variables: the distance between two categories is computed as a function of their co-occurrence with other categories. A frequency-based representation of the cluster centroid is used for categorical variables, whereas the mean is used for continuous variables. Another well-known method for clustering mixed-type data is K-prototypes (Huang 1998), a variant of K-means that is based on the weighted combination of the squared Euclidean distance for continuous variables and the matching distance for categorical variables. A major concern when using any of the aforementioned techniques is how to choose a suitable weight for each variable or variables of the same type (Hennig and Liao 2013; van de Velden et al. 2019). For instance, Gower's similarity measure assigns equal weights to both categorical and continuous variables in computing the similarity between two observations.

Since most hierarchical clustering methods are designed to deal with variables measured on the same scale, another popular strategy is to homogenize the variables in a preprocessing phase. For mixed data, this would mean re-coding the continuous variables as categorical ones or vice-versa and then use a (dis)similarity measure suitable for continuous or categorical data, respectively. Typical discretization of continuous variables presents two major drawbacks: (a) substantially different values can be put in the same class or interval, which implies loss of information, and (b) values that are close to each other, that is on the boundary of two neighboring classes, can be put in different classes, leading to information distortion. An alternative transformation through discretization is to code the original values of a continuous variable into a prespecified number of fuzzy categories, that is, to a set of $k$ nonnegative values that sum to 1, quantifying the "possibility" of the variable to be in each category (van Rijckevorsel 1988; Aşan and Greenacre 2011). Rather than cut-points, $h$ membership functions can be used, with $h$ "hinge points" or pivots. They may have different shapes such as triangular, trapezoidal, Gaussian, and so forth. For instance, in the case of the so-called triangular membership functions with $h = 3$, the hinge points could be the minimum, the median and the maximum. A similar scheme has been proposed by Escofier (1979), where a standardized variable $z$ (with mean 0 and variance 1) can be expressed as two fuzzy variables, $y_+ = (1+z)/2$ and $y_- = (1-z)/2$. This "bipolar" coding reflects to what extent an observation lies above or below the mean. The general principle of such coding schemes is that a constant unit is spread across two or more categories of the recoded variable.

In this work, we present a coding scheme that enables hierarchical clustering with categorical (nominal or ordinal) and continuous (interval) variables. Each categorical variable is replaced by a set of dummy (0-1) variables, that is, a separate column is created for each category and observed categories are coded using ones, whereas all other observations are zeros. Each continuous variable is recoded via a generalization of the 0-1 coding, carried out in two steps. In the first step, each continuous variable is transformed into an $m$-point ordinal scale, where the choice of $m$ is data-driven and depends on the variable's range. In the second step, each value of the derived ordinal scale is recoded into an $n$-tuple of nonnegative values that add up to 1. These can be considered probabilities that an observation lies in the respective category of the ordinal scale. The value of $n$ is user-specified and corresponds to the number of fuzzy-coded variables that are created. With regard to existing approaches, the result of the second step is similar to fuzzy coding (van Rijckevorsel 1988; Aşan and Greenacre 2011), where zeros and ones are replaced by fractions to avoid information loss related to discretization. However, instead of using membership functions, our proposal is based on the concept that each value of an ordinal variable is the midpoint of an interval and carries a mass of 1 that is decomposed in a barycentric fashion into intervals of subsequent or previous values.

As it will be shown in Section 2, the proposed recoding scheme is accurate and sensitive even at the smallest differences between the values of a continuous variable, minimizing information loss. Finally, the proposed scheme can be paired with a Ward hierarchical clustering (see, e.g., Murtagh and Legendre 2014) using the chi-square distance, very much in the spirit of (Multiple) Correspondence Analysis.

The outline of the paper is as follows. Section 2 presents the proposed barycentric coding scheme for continuous variables. The recoded data is given as input to a hierarchical clustering algorithm based on Ward's linkage criterion and the chi-square distance. This is discussed in Section 3. In Section 4.1, the proposed approach and four alternative clustering approaches for mixed-type data are illustrated on three real-world data sets with different numbers of continuous and categorical variables. A simulation study in Section 4.2, examines the effectiveness of the proposed approach on simulated data under varying number of clusters, sample size, cluster density, cluster overlap and numbers of continuous and categorical variables. The paper concludes in Section 5.

## 2 A barycentric coding of continuous variables

To facilitate understanding of the proposed approach, in Table 1, we first consider a toy example of the recoding of a continuous variable, X, with seven values, ranging from 32 to 60.1. In Step 1, the continuous variable is recoded into a 57-point ordinal scale, $X_{ord}$. In the second step, the ordinal scale derived from the previous step, is recoded into a set of $n$-tuples of nonnegative values that add up to 1. Table 1 shows the final result for $n$ = 3 (columns $X_1$ to $X_3$) and $n = 5$ (columns $X_1$ to $X_5$). The two steps of the proposed scheme are thoroughly described below.

Table 1. The results of barycentric recoding of a continuous variable, X, into three ordered categories (or 3-tuples) and five ordered categories (or 5-tuples)

| Step 1 | | Step 2 ($n$ = 3) | | | Step 2 ($n$ = 5) | | | | |
|---|---|---|---|---|---|---|---|---|---|
| X | $X_{ord}$ | $X_1$ | $X_2$ | $X_3$ | $X_1$ | $X_2$ | $X_3$ | $X_4$ | $X_5$ |
| 40.0 | 17 | 0.711 | 0.193 | 0.096 | 0.184 | 0.667 | 0.099 | 0.033 | 0.017 |
| 33.0 | 3 | 0.956 | 0.029 | 0.015 | 0.927 | 0.049 | 0.016 | 0.005 | 0.003 |
| 32.5 | 2 | 0.974 | 0.018 | 0.009 | 0.956 | 0.029 | 0.010 | 0.003 | 0.002 |
| 32.0 | 1 | 0.991 | 0.006 | 0.003 | 0.985 | 0.010 | 0.003 | 0.001 | 5E-04 |
| 55.2 | 47 | 0.061 | 0.123 | 0.816 | 0.011 | 0.023 | 0.068 | 0.205 | 0.693 |
| 60.1 | 57 | 0.003 | 0.006 | 0.991 | 5E-04 | 0.001 | 0.003 | 0.010 | 0.985 |
| 32.0 | 1 | 0.991 | 0.006 | 0.003 | 0.985 | 0.01 | 0.003 | 0.001 | 5E-04 |

*Step 1:* Transformation of a continuous variable into an $m$-point ordinal scale.

Let $\mu$ be the minimum and $M$ the maximum observed values of a continuous variable, X.

1. Calculate the smallest distance, $d_0$, between all successive observed values of X.

2. Find the unique integer, $m$, for which it holds $m - 1 \leq \frac{M-\mu}{d_0} \leq m$.

3. Divide the closed interval [$\mu$, $M$] of observed values into $m$ intervals of equal width, $d_0$, which are half-open intervals of the form:

$[\mu, \mu + d_0), [\mu + d_0, \mu + 2d_0), \dots, [\mu + (\mu - 2)d_0, \mu + (m - 1)d_0), [\mu + (m - 1)d_0, \mu + md_0].$

It is trivial to show that each interval $[\mu + id_0, \mu + (i + 1)d_0]$, contains no more than a single value of X and the last interval contains the maximum, $M$, since $\mu + (m - 1)d_0 \leq M \leq \mu + md_0$.

4. The minimum, $\mu$, that belongs to the interval $[\mu, \mu + d_0)$ and all values equal to the minimum are assigned to value 1. In general, each observed value, $x$, is assigned to an integer $T(x) = \left[\frac{x-\mu}{d_0}\right] + 1$, from 1 to $m$, where $[a]$ indicates the known integer part of a real number $\alpha$. The maximum, $M$, is assigned to value $m$.

Based on this transformation, each observed value is assigned to only one of the integer values $1, 2, \dots, m$ of an $m$-point ordinal scale, where two successive values, even if they are quite close to each other, are assigned to different values of the ordinal scale. Note that some of the values $1, 2, \dots, m$ may not be matched to any of the initial values. Also, the smaller the distance, $d_0$, is, the larger the number of intervals, $m$, in which the continuous variable, X, is divided, and the larger the corresponding number of points of the ordinal scale. Finally, note that each value $x$ of the continuous variable, X, belongs to one interval only and therefore corresponds to a unique point of the $m$-point scale.

The domain of the transformation, $f$, is the set of all intervals where the observed values of X fall into, and the image is all numbers $1, 2, \dots, m$, of the $m$-point scale. Obviously, transformation $f$ is one-to-one, hence reversible. The inverse transformation, $f^{-1}$, maps an integer of the $m$-point scale, from those assigned via the transformation, $T(x)$, to the thin interval of width $d_0$ that contains an observed value, $x$. Given the minimum, $\mu$, the maximum, $M$ and the integer $m$, it is trivial to show that the smallest distance, $d_0$, satisfies the condition $\frac{M-\mu}{m} < d_0 \le \frac{M-\mu}{m-1}$. Therefore, the original intervals that contain the values of X can be easily determined; this can lead to the reconstruction of the observed values with a minimal loss of information.

*Example.* Given the seven values of the continuous variable, X, in Table 1, with $\mu = 32$ and $M$ = 60.1, the minimum distance, $d_0 = 0.5\ (32.5 - 32)$. Therefore, $\frac{M-\mu}{d_0} = 56.2$, the total number of points, $m = 57$, and each value of X will be assigned into one of the integer values $1, 2, 3, \ldots, 57$. The first value, 40, corresponds to the integer, $T(40) = \left[40 - \frac{32}{0.5}\right] + 1 = [16] + 1 = 17$. Therefore, 40 is recoded into 17 in the 57-point scale. Similarly, 33 is recoded into 3, which belongs to the interval [33, 33.5), i.e., the third interval etc.

*Step 2.* Recoding of an *m*-point ordinal scale to an *n*-tuple of $\mathbb{R}^n$.

In this step, the values of the *m*-point ordinal scale are recoded into *n*-tuples of the vector space $\mathbb{R}^n$ via the barycentric coding scheme of Moschidis and Chadjipadelis (2017). The integer, *n*, is user-specified and denotes the number of fuzzy-coded variables (new columns) that will be created. The values of the *n*-tuples add up to 1 (see Table 1). The rationale behind this recoding scheme is that each point of the *m*-point ordinal scale carries a weight (or mass) equal to 1. This mass is then distributed in a recursive manner into *n* parts, that correspond to the *n* values of each *n*-tuple. Since this process is crucial in the proposed clustering method, it is thoroughly described below.

Let $l \in \{1, 2, \ldots, m\}$ a value of an *m*-point ordinal scale (e.g., any value of $X_{ord}$ in Table 1) and $y_{l1}, y_{l2}, \ldots, y_{ln}$ the corresponding values of an *n*-tuple of $\mathbb{R}^n$ that add up to 1. The interval $\left(\frac{1}{2}, m + \frac{1}{2}\right)$ can be divided into $n$ equal intervals $\left(\frac{1}{2}, \frac{1}{2} + \frac{m}{n}\right), \left[\frac{1}{2} + \frac{m}{n}, \frac{1}{2} + 2\frac{m}{n}\right), \ldots, \left[\frac{1}{2}(n-1)\frac{m}{n}, \frac{1}{2} + n\frac{m}{n}\right)$, respectively. Let $B_i = \frac{1}{2} + i\frac{m}{n}$, for $i = 0, 1, \ldots, n$ the $n + 1$ bounds of those intervals and $A_i = \left(\frac{1}{2} + (i-1)\frac{m}{n}\right) + \frac{m}{2n}$, for $i = 1, 2, \dots, n$ their corresponding midpoints. Following the example given in Table 1, for $m = 57$ and $n = 3$, the interval $(B_0, B_3) = \left(\frac{1}{2}, m + \frac{1}{2}\right) = \left(\frac{1}{2}, 57 + \frac{1}{2}\right)$ is divided into $(B_0, B_1) = \left(\frac{1}{2}, \frac{1}{2} + \frac{57}{3}\right) = (0.5, 19.5)$, $[B_1, B_2) = \left[\frac{1}{2} + \frac{57}{3}, \frac{1}{2} + 2\frac{57}{3}\right) = [19.5, 38.5)$ and $[B_2, B_3) = \left[\frac{1}{2} + 2\frac{57}{3}, \frac{1}{2} + 3\frac{57}{3}\right) = [38.5, 57.5)$, respectively. Then, $A_1 = 10$, $A_2 = 29$ and $A_3 = 48$ are the corresponding interval midpoints.

Our aim, as stated above, is to recode every value, $l$, of an $m$-point scale to an $n$-tuple of $\mathbb{R}^n$, with a sum equal to 1. This sum (or mass) is recursively decomposed into $n$ parts, which are attributed to the $n$ values of the $n$-tuple in question. The decomposition process is presented below for a given $l$.

It is well-known that if $g$ is the center of gravity of points $x_1$ and $x_2$ with masses $m_1$ and $m_2$, respectively, then:

$$g = \frac{m_1 x_1 + m_2 x_2}{m_1 + m_2} \quad (1)$$

and $q = m_1 + m_2$, the total mass. Assume that a total mass $q = 1$ is associated with each point $l$ of the $m$-point ordinal scale. For $g = l$, we have

$$m_1 = \frac{x_2 - g}{x_2 - x_1} q = \frac{x_2 - l}{x_2 - x_1} \quad (2).$$

If we now consider the decomposition of this mass into an $n$-tuple of $\mathbb{R}^n$, we can distinguish two cases:

*A*. Value $l$ lies in the first or the last interval.

Value $l$ lies in the first out of the $n$ intervals, i.e. the interval $(B_0, B_1)$. Let's assume that $l$ is the center of gravity of points $x_1 = \frac{1}{2} = B_0$ and $x_2 = \frac{1}{2} + \frac{m}{n} = B_1$, respectively, with masses $m_1$ and $m_2$. From Eq. (2) we have,

$$m_1 = \frac{\left(\frac{1}{2} + \frac{m}{n}\right) - l}{\left(\frac{1}{2} + \frac{m}{n}\right) - \frac{1}{2}} = \frac{\frac{1}{2} - l + \frac{m}{n}}{\frac{m}{n}} = 1 + \left(\frac{1}{2} - l\right)\frac{n}{m}.$$

Consequently, $m_2 = 1 - m_1 = 1 - \left(1 + \left(\frac{1}{2} - l\right)\frac{n}{m}\right) = \left(l - \frac{1}{2}\right)\frac{n}{m}$.

The mass $m_2 = \left(l - \frac{1}{2}\right)\frac{n}{m}$ at point $B_1$, further decomposes into $m_1'$ at $A_1$, the midpoint of $B_0$ and $B_1$, and $m_2 - m_1'$ at $B_2$. Since $A_1$ is the midpoint of the first interval, $B_1$ is the upper bound of the first interval, and $B_2$ the upper bound of the second interval, the ratio of the distances from $B_2$ to $B_1$ and from $B_2$ to $A_1$ is always equal to $\frac{2}{3}$. Thus,

$$m_1' = \left(\frac{B_2 - B_1}{B_2 - A_1}\right) m_2 = \frac{2}{3} m_2 = \frac{2}{3}\left(l - \frac{1}{2}\right)\frac{n}{m}.$$

Then the first element, $y_{l1}$, of the $n$-tuple is given by:

$$y_{l1} = m_1 + m_1' = 1 + \left(\frac{1}{2} - l\right)\frac{n}{m} + \frac{2}{3}\left(l - \frac{1}{2}\right)\frac{n}{m} = 1 + \frac{1}{3}\left(\frac{1}{2} - l\right)\frac{n}{m}.$$

The remaining mass at $B_2$ is equal to $\frac{1}{3} m_2$ and is decomposed at points $B_3$ and $A_2$. Since $A_2$ is the midpoint of the second interval and $B_3$ is the upper bound of the third interval, the ratio of the distances from $B_2$ to $B_3$ and from $B_2$ to $A_2$ and is $\frac{2}{3}$. Therefore,

$$y_{l2} = \frac{2}{3}\left(\frac{1}{3} m_2\right) = \frac{2}{3^2}\left(l - \frac{1}{2}\right)\frac{n}{m}$$

Similarly, the remaining mass $\frac{1}{3}\left(\frac{1}{3} m_2\right)$ of point $B_3$, is decomposed into two parts, at points $B_4$ and $A_3$. Repeating the same process, we have $y_{li} = \frac{2}{3^i}\left(l - \frac{1}{2}\right)\frac{n}{m}$, for $i = 2, 3, \ldots, n-1$. The last element, $y_{ln}$, is given by:

$$y_{ln} = \frac{1}{3}\frac{1}{3^{n-2}}(l-1)\frac{n}{m} = \frac{1}{3^{n-1}}\left(l - \frac{1}{2}\right)\frac{n}{m}.$$

*Remark.* If $l$ belongs to the last interval, it is trivial to show that, due to symmetry, the corresponding $n$-tuple is equal to $y_{ln}, y_{l(n-1)}, \ldots, y_{l1}$ and its elements can be computed starting from the last interval.

Referring to the example given in Table 1, for $l = 3$ we derive the values $y_{31} = 0.956$, $y_{32} = 0.029$ and $y_{33} = 0.015$, as follows: $l = 3$ lies in the first interval $(B_0, B_1) =$ $(0.5, 19.5)$. Then, $m_1 = 1 + \left(\frac{1}{2} - l\right)\frac{n}{m}$ =1+$\left(\frac{1}{2} - 3\right)\frac{3}{57} = 0.8684$ and $m_1' = \frac{2}{3}\, m_2 = \frac{2}{3}(1 - m_1) = \frac{2}{3}(1 - 0.8684) = 0.088, y_{31} = m_1 + m_1' = 0.8684 + 0.088 = 0.956,$

$y_{32} = \frac{2}{3^2}\left(l - \frac{1}{2}\right)\frac{n}{m} = \frac{2}{3^2}\left(3 - \frac{1}{2}\right)\frac{3}{57} = 0.029$ and $y_{33} = \frac{1}{3^2}\left(l - \frac{1}{2}\right)\frac{n}{m} = \frac{1}{3^2}\left(3 - \frac{1}{2}\right)\frac{3}{57} = 0.015.$

Notice that the coded data can be transformed back to the original (ordinal) data. For instance, given that $y_{31} = 0.956$ we find that $l = 3$ satisfies $1 + \frac{1}{3}\left(\frac{1}{2} - l\right)\frac{3}{57} = 0.956$.

B. Value *l* lies in a middle interval.

Let's assume that *l* lies in the $p + 1$ interval. Then it holds:

$$B_p = \frac{1}{2} + p\frac{m}{n} < l < \frac{1}{2} + (p + 1)\frac{n}{m} = B_{p+1}.$$

Given that *l* is the center of gravity, the total mass is decomposed into $m_1$ and $m_2$ at points $B_p$ and $B_{p+1}$, respectively. From Eq. (2), we have:

$$m_1 = \frac{\left[\frac{1}{2} + (p + 1)\frac{m}{n}\right] - l}{\left[\frac{1}{2} + (p + 1)\frac{m}{n}\right]\left(\frac{1}{2} + p\frac{m}{n}\right)} = \left[\left(\frac{1}{2} + (p + 1)\frac{m}{n}\right) - l\right]\frac{n}{m} = \left(\frac{1}{2} - l\right)\frac{n}{m} + (p + 1),$$

$$m_2 = 1 - m_1 = \left(l - \frac{1}{2}\right)\frac{n}{m} - p.$$

The mass, $m_1$, at point $B_p$ is further decomposed into $\frac{1}{3}\, m_1$ at point $B_{p-1}$ and $\frac{2}{3}\, m_1$ at point $A_{p+1}$. Similarly, the mass, $m_2$, at point $B_{p+1}$ decomposes into $\frac{2}{3}\, m_2$ at $A_{p+1}$ and $\frac{1}{3}\, m_2$ at $B_{p+2}$. The total mass at point $A_{p+1}$ (the midpoint of the $p + 1$ interval) is equal to $\frac{2}{3}\, m_1 + \frac{2}{3}\, m_2 = \frac{2}{3}(\,m_1 + m_2) = \frac{2}{3}$. This quantity is the $p + 1$ element of the corresponding *n*-tuple, i.e., $y_{l(p+1)} = \frac{2}{3}$. The remaining elements can be calculated as described in case A, where *l* belongs to the first interval. For example, the element $y_{l(p+i)}$ is given by $\frac{2}{3^{i-1}}\left(\frac{1}{3}\, m_2\right)$, while the element $y_{ln}$ is given by $\frac{1}{3^{n-1}}\left[\left(l - \frac{1}{2}\right)\frac{n}{m} - p\right]$.

It is trivial to verify that for a standardized variable (with mean 0 and variance 1), the results in both steps of the algorithm, i.e., the derived ordinal scale and the $n$-tuples, are identical to those in the unstandardized case. Note also that Step 2 can be used for recoding ordinal variables that are Likert-type scales or $m$-point rating scales. In that case, Step 1 is skipped and Step 2 can be applied directly to the ordinal scale, allowing to maintain ordinality.

## 3 Hierarchical clustering of mixed-type data

In this section, we present a hierarchical clustering algorithm based on Ward's linkage criterion and the chi-square distance. We illustrate that this algorithm is compatible with the proposed recoding scheme and can be subsequently applied to the transformed data matrix.

Let $\mathbf{Z}$, the $I \times J$ matrix obtained from the barycentric coding of the continuous variables and the dummy (0-1) coding of the categorical variables. Notice that the sum of all the elements in a row of $\mathbf{Z}$ is equal to the number of variables. Because the matrix $\mathbf{Z}$ can be considered as a kind of contingency table, the chi-square distance can be used to determine the similarity between rows. The chi-square distance is the most well understood and defined metric for non-quantitative data and typically used for contingency and categorical data (Greenacre and Hastie 1987); it is also the basis of Correspondence Analysis (CA) (Greenacre 2017). There exists a theoretical justification for the choice of this particular distance function for categorical data, as the chi-square distance is a type of Mahalanobis distance (Greenacre 2017, p. 299).

In the context of CA, the $I \times J$ matrix of relative frequencies, $\mathbf{P}$, known as the *correspondence matrix*, is obtained by dividing the frequencies in $\mathbf{Z}$ by the overall total frequency, i.e., $n = \sum_{ij} z_{ij}$, where $z_{ij}$ indicates the $ij$-th element of the matrix $\mathbf{Z}$. The total frequency, $n$, is equal to the sample size. The row marginals of $\mathbf{P}$ form the vector of the row masses, $\mathbf{r}$, with elements $r_i = \sum_{j=1}^{J} p_{ij}$. The column marginals of $\mathbf{P}$ form the vector of the column masses, $\mathbf{c}$, with elements $c_j = \sum_{i=1}^{I} p_{ij}$. The so-called row profiles, $\mathbf{a}_1, \mathbf{a}_2, \dots, \mathbf{a}_I$ are vectors of length $J$ and are obtained by dividing each row of $\mathbf{P}$ by its marginal total. The chi-square distance is defined as the distance between the $i$-th row profile, $\mathbf{a}_i$ and the $j$-th row profile, $\mathbf{a}_j$, as follows:

$$d_{\chi^2}(\mathbf{a}_i, \mathbf{a}_j) = (\mathbf{a}_i - \mathbf{a}_j)^{\mathrm{T}} \mathbf{D}_{1/c} (\mathbf{a}_i - \mathbf{a}_j),$$

where $\mathbf{D}_{1/c}$ is a diagonal matrix with elements the inverse of the column masses, $\mathbf{c}$.

The chi-square distance incorporates a weight that is inversely proportional to the total of each column, which increases the importance of small deviations in the columns which have a small weight with respect to those with a large weight. Note also that the chi-square distance between observations on the recoded variables is equal to the Euclidean distance, when all columns have equal weights or masses (Moschidis 2015; Moschidis and Chadjipantelis 2017). Furthermore, the chi-square distance has the property of "distributional equivalence" (Greenacre 2017, p.37), meaning that it ensures that the distances between rows are invariant when two columns are aggregated. It is thanks to this property that the exact choice of $n$ in Step 2 of barycentric coding is not so critical. Finally, an equal number of categories, $n$, can be created for each continuous variable, so that the chi-square distances between observations are not affected by the different number of categories, as it has been reported elsewhere (Le Roux and Rouanet 2004). This leads to a homogenization of continuous variables.

Based on the chi-square distance, a special case of agglomerative Ward's minimum variance method can be defined (see, Greenacre 2017, p.120). In this type of clustering, groups are merged according to a minimum-distance criterion which takes into account the weights of each observation (row) being clustered. The specific (squared) "distance" between two row groups $g$ and $h$ is given by:

$$\frac{\bar{r}_g \bar{r}_h}{\bar{r}_g + \bar{r}_h} d_{\chi^2}^2(\bar{\mathbf{a}}_g, \bar{\mathbf{a}}_h),$$

where $\bar{r}_g$ and $\bar{r}_h$ are the masses of the respective groups (calculated as the sum of the masses of the members of each group), $\bar{\mathbf{a}}_g$ and $\bar{\mathbf{a}}_h$ are the profiles of the groups (that is, the profiles of the merged rows) and $d^2_{\chi^2}(\bar{\mathbf{a}}_g, \bar{\mathbf{a}}_h)$ denotes the squared chi-square distance between the profiles of the groups.

At each step, Ward's algorithm consists in aggregating two groups such that the increase of within-cluster variance (or inertia) is minimum; this is equivalent to minimizing the reduction of between-cluster inertia (Lê et al., 2008). This procedure is represented by a hierarchical tree, which is considered as a sequence of nested partitions and provides a decomposition of inertia, analogous to the decomposition of inertia with respect to principal components in the context of Correspondence Analysis. The number of clusters can then be chosen by looking at the bar plot of the changes in inertia (or inertia gains). An empirical criterion suggests a division into $K$ clusters when the increase of between-cluster inertia between $K – 1$ and $K$ clusters is much greater than the one between $K$ and $K + 1$ clusters. Let $\Delta(K)$ denote the between-cluster inertia increase when moving from $K – 1$ to $K$ clusters. The selected number of clusters, $K$, is the one that minimizes the ratio $\Delta(K)/\Delta(K+1)$ (Lê et al., 2008).

## 4 Applications

### 4.1 Real data

In this section, we apply the proposed approach and alternative methods for clustering mixed-type data to a series of real data sets with known clustering structure, in order to assess the agreement between clustering partitions obtained with different methods. Five methods were considered: #1 Gower's dissimilarity measure followed by Partitioning Around Medoids (PAM; Kaufman and Rousseeuw 1990) on the corresponding distance matrix, #2 K-prototypes (Huang, 1998), #3 mixed K-means (Ahmad and Dey 2007), #4 Ward's hierarchical clustering based on fuzzy coding with triangular membership functions (Aşan and Greenacre 2011) and

#5 Ward's hierarchical clustering based on barycentric coding (proposed approach). For these methods, software implementations are readily available in R. In particular, we used the fuzzy.tri() function for fuzzy coding of continuous variables (Greenacre 2013), available online at http://www.esapubs.org/archive/ecol/E094/021/fuzzy.tri.R and the packages cluster (Maechler et al. 2018) for Gower's dissimilarity followed by PAM**, clustMixType** (Szepannek 2017) for K-prototypes and DisimForMixed (Pathberiya 2016) for mixed K-means. The proposed approach is implemented in R; the source code of barycentric coding can be found in the Appendix.

Three real-world data sets, characterized by a different proportion of continuous and categorical variables, were considered: a data set with more categorical than continuous variables (Cleveland heart disease), one with more continuous than categorical variables (credit approval data set) and a balanced scenario (diamond stone price data set). The three data sets are presented below. The adjusted Rand index (ARI) (Hubert and Arabie 1985) was used to assess the level of agreement between clustering partitions obtained with different methods, as well as their agreement with an a priori known clustering structure or an informative external criterion, where available. The ARI assumes a value of unity when there is perfect recovery of the true cluster structure and a value of zero when recovery is equal to random chance.

*Cleveland Heart Disease.* The Cleveland Heart Disease data set contains information concerning heart disease diagnosis. The data was collected from the Cleveland Clinic Foundation, and it is available at the UCI machine learning Repository (https://archive.ics.uci.edu/ml/datasets/Heart+Disease). It consists of five quantitative and eight categorical variables measured on 303 patients. Following earlier studies that used this data set (Hunt and Jorgensen 2011; van de Velden et al. 2019), the heart disease variable was recoded in two categories denoting the presence or absence of heart disease. This variable was used as an external (true) partition to evaluate the obtained clustering solutions. Six observations that had missing values in any of the 13 variables analyzed were omitted from further analysis, leaving 297 observations.

*Credit Approval.* The credit approval data have 653 complete observations assigned to two classes. The class distribution is 303 and 307, respectively. The data set consists of six numerical and nine categorical variables and was also taken from the UCI Machine Learning Repository (https://archive.ics.uci.edu/ml/datasets/credit+approval).

*Diamond stone pricing.* The data set is available from the JSE Data Archive (http://jse.amstat.org/jse_data_archive.htm) and refers to the characteristics that influence the price of diamond stones ($n$ = 308). It contains one nominal variable (certifying agency), two ordinal variables (colour, clarity), and three quantitative ones (small size, medium size, large size). Following Chu (2001) and Vichi et al. (2019), the three quantitative variables have been derived from three binary variables that represented the three different sizes of diamonds, small (less than 0.5 carats), medium (0.5 to less than 1 carat) and large (1 carat and over). The three binary variables were multiplied by carats to derive the three quantitative variables. An additional variable, diamond stone price in $, has been left out of the analysis. Previous analyses of this data set (Vichi et al. 2019) showed that three clusters exhibit a good separation in discriminating diamond prices.

Table 2 shows the degree of agreement between methods in terms of ARI in the three datasets. Our approach, Ward’s hierarchical clustering based on barycentric coding, is denoted as “mixed hierarchical B”, whereas hierarchical clustering based on fuzzy coding using triangular membership functions is denoted as “mixed hierarchical T”. In all cases, continuous variables were recoded into two categories, but similar results were obtained with three, four and five categories.

Table 2. Agreement between methods (ARI values) for each of the considered mixed-type data sets; from left to right, Cleveland heart disease, credit approval and diamond stone pricing data sets.

| **Data set** | **Heart ($K$ = 2)** | | | | | **Credit ($K$ = 2)** | | | | | **Diamond ($K$ = 3)** | | | | |
|---|---|---|---|---|---|---|---|---|---|---|---|---|---|---|---|
| **Method** | **(1)** | **(2)** | **(3)** | **(4)** | **(5)** | **(1)** | **(2)** | **(3)** | **(4)** | **(5)** | **(1)** | **(2)** | **(3)** | **(4)** | **(5)** |
| **(1) Gower/PAM** | - | | | | | - | | | | | - | | | | |
| **(2) K-prototypes** | .36 | - | | | | .66 | - | | | | .46 | - | | | |
| **(3) Mixed K-means** | .20 | .36 | - | | | .11 | .25 | - | | | .46 | 1 | - | | |
| **(4) Mixed hierarchical T** | .45 | .27 | .23 | - | | .00 | .00 | .01 | - | | .45 | .57 | .57 | - | |
| **(5) Mixed hierarchical B** | .53 | .29 | .24 | .77 | - | .00 | .00 | .04 | .91 | - | .46 | 1 | 1 | .57 | - |

For the diamond stone pricing data set, where $K$ was set to 3, based on previous studies (Vichi et al. 2019), mixed hierarchical B, K-prototypes and mixed K-means produced identical partitions (ARI = 1). Figure 1 shows the dendrogram generated using mixed hierarchical B and a barplot of the inertia gains. The height of the fusion, provided on the vertical axis of the dendrogram, indicates the (dis)similarity/distance between two clusters/observations and corresponds to inertia. The upper limit of these rectangles indicates the point at which the tree has been cut. Notice that, based on the change in inertia, the optimal number of clusters is three. The three clusters are shown in rectangles and their content is consistent with previous analyses of the data; it exhibits a good separation in discriminating the values of diamond prices. The first cluster (C1 in Fig. 1, 30.5%), contains small size diamond stones (less than 0.5 carats) with IGI certification and lower than average price. The second cluster (C2 in Fig. 1, 23%), consists of large stones (more than 0.5 carats) with HRD certification and higher price than average, and the third cluster (C3 in Fig. 1, 46.4%) contains medium size stones (around 0.5 carats) with HRD certification and moderate price. Gower/PAM and mixed hierarchical T, however, failed to discriminate between diamond stones with low and moderate prices.

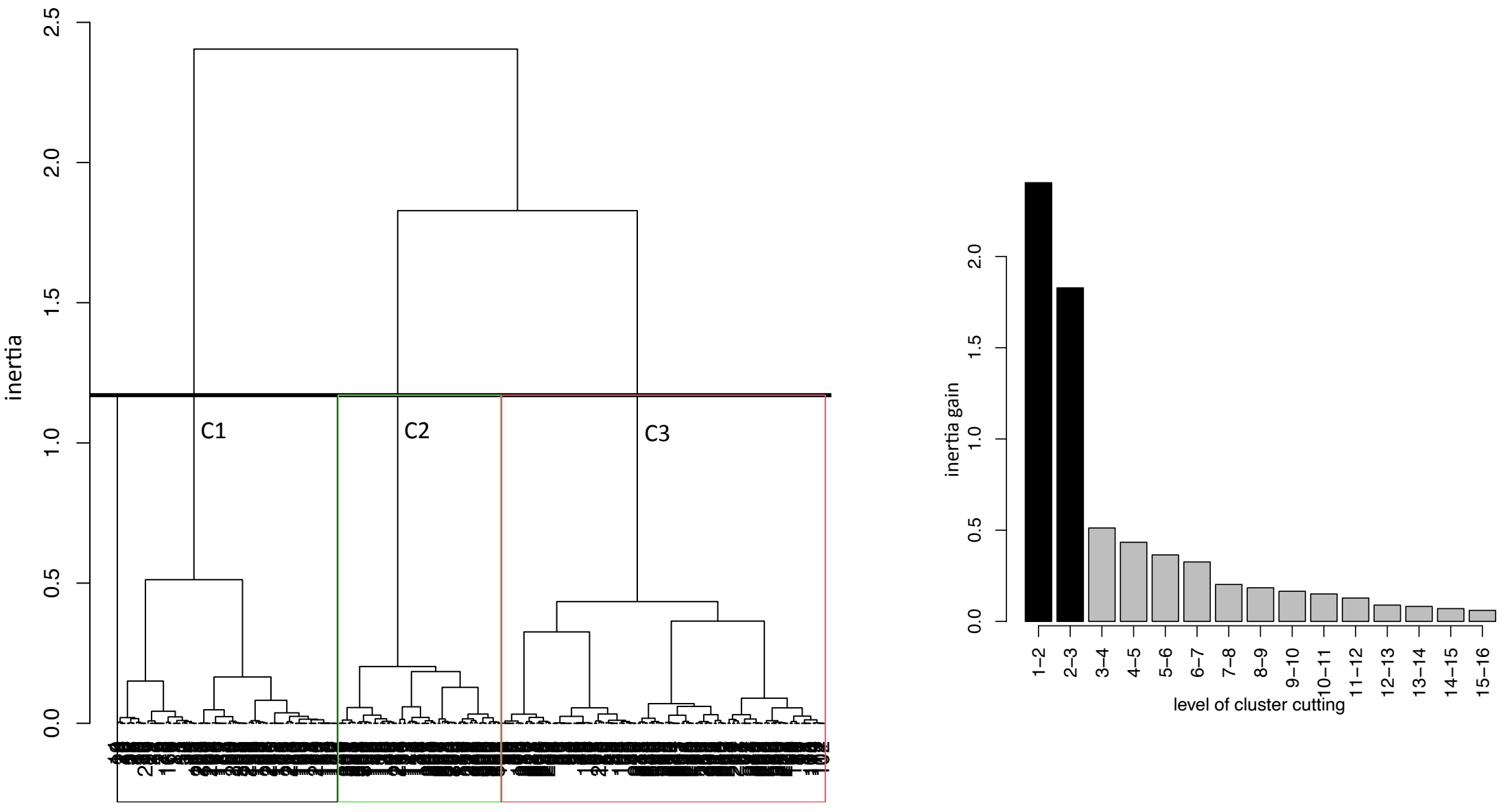

Figure 1. Dendrogram of the mixed hierarchical clustering based on barycentric coding on the diamond stone pricing data, leading to three clusters, C1 to C3 (left) and barplot of inertia gains (right).

For the Cleveland Heart Disease data set, where the number of clusters, *K*, is a priori known and set to 2, mixed hierarchical approaches and Gower/PAM showed the highest agreement with the true partition (.39 and .38, respectively). As expected, the hierarchical methods resulted in the most similar partitions (.77), followed by a moderate degree of agreement between hierarchical methods and Gower/PAM (.53 for barycentric and .45 for triangular coding, respectively). Figure 2 shows the dendrogram generated using mixed hierarchical B and a barplot of the inertia gains. The latter indicates that the optimal number of clusters is two. The first cluster (C1 in Fig. 2, 43.8%) contains 80.3% of the patients with heart disease. Patients in this cluster are mostly males with asymptotic chest pain, exercise-induced angina, a reversible defect in thalassemia and flat ST segment depression (where ST segment is an isoelectric section of the electrocardiogram). They also have higher than average resting blood pressure and old peak (pressure of the ST segment). The second cluster (C2 in Fig. 2, 56.2%) contains 81.2% of the patients without heart disease. These are mostly females without chest pain or exercise-induced angina, no thalassemia, lower than average resting blood pressuring and upsloping ST-segment depression.

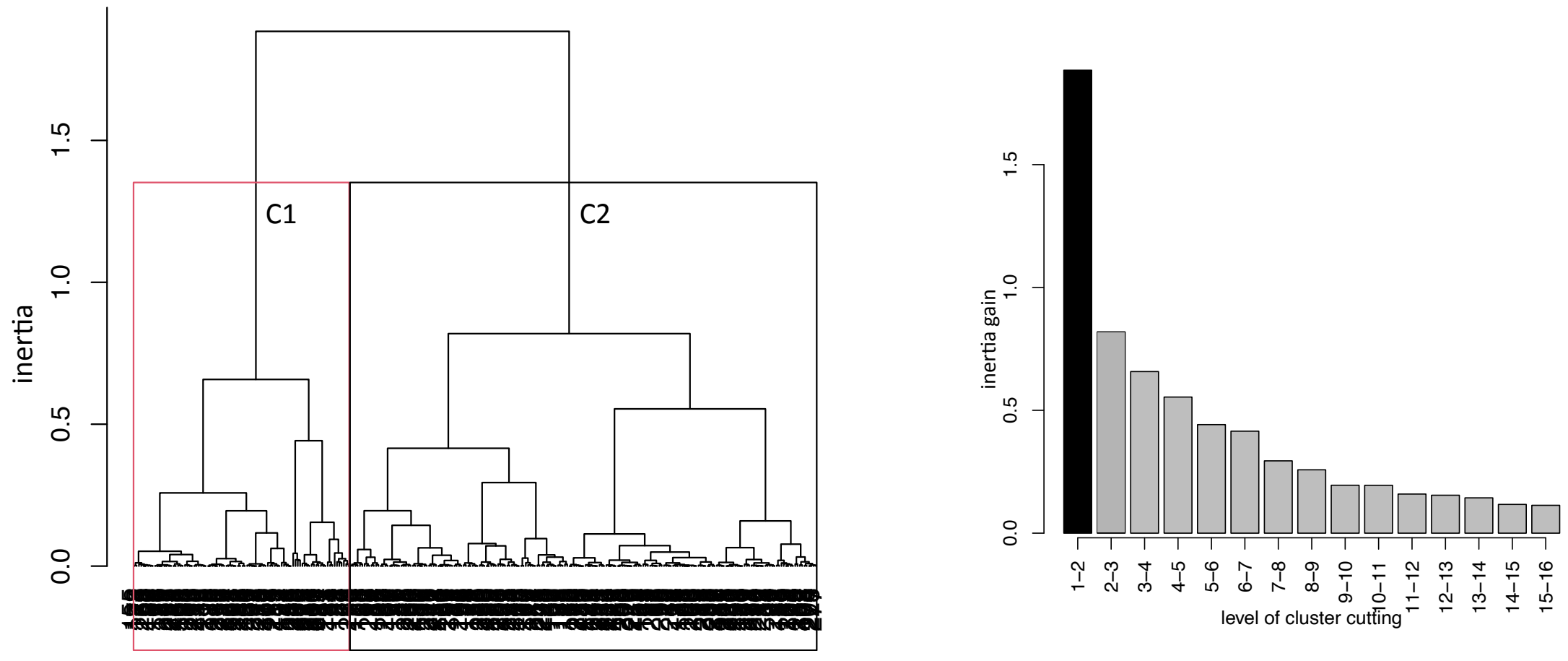

Figure 2. Dendrogram of the mixed hierarchical clustering based on barycentric coding on the Cleveland heart disease data, leading to two clusters, C1 and C2 (left) and barplot of inertia gains (right).

Last, for the credit approval data set (with the number of clusters set to two), hierarchical methods resulted in the most similar partitions (.91). These partitions, however, had nothing in common with the ones obtained from the other three methods (ARI values near zero). Gower/PAM and K-prototypes also had a high degree of agreement (.66). This can be explained due to the fact that the solutions of both hierarchical methods did not agree at all with the true clustering structure (ARI < .10), whereas Gower/PAM and K-prototypes performed much better (ARI values of .38 and .37, respectively). The dendrogram of mixed hierarchical B and the barplot of the inertia gains are shown in Figure 3. The optimal number of clusters is six, but only two clusters were considered for comparison purposes. Notice also that since the variable names and their values are meaningless to protect the confidentiality of the data, interpretation of the clusters is not obvious. The first cluster (C1 in Fig. 3, 8.1%) is characterized mainly by higher-than-average values of variable 2 and the presence of the categories ‘ff’ in variables 6 and 7. The second cluster (C2 in Fig. 3, 91.8%) is characterized mainly by lower-than-average values of variable 2 and the presence of the categories ‘v’ and ‘h’ of variable 7, ‘+’ of variable 16 and ‘c’ of variable 6.

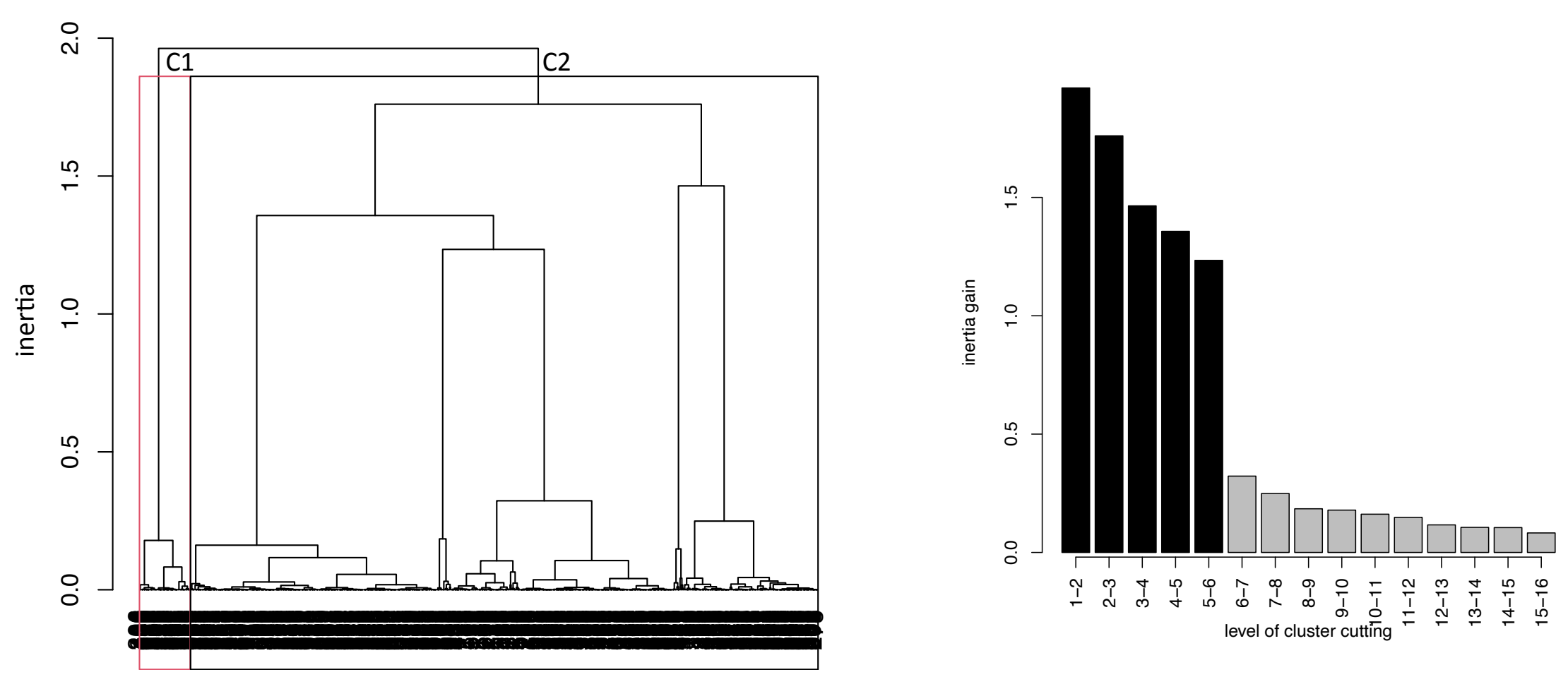

Figure 3. Dendrogram of the mixed hierarchical clustering based on barycentric coding on the credit approval data, leading to two clusters, C1 and C2 (left) and barplot of inertia gains (right).

### 4.2 Simulation study

4.2.1 Data generation

A simulation study was conducted to determine the effectiveness of each of the five alternative procedures considered in Section 4.1. The data were generated using the R package MixSim (Melnykov, Chen, & Maitra 2012), which allows for the determination of pairwise overlap between any pair of clusters. The notion of pairwise overlap is defined as the sum of two misclassification probabilities for a pair of weighted Gaussian distributions (Maitra and Melnykov 2010). However, this notion is only determined for continuous variables and in order to obtain a categorical variable for the simulation data, a continuous variable was discretized by dividing it into *c* classes with the 100/*c*% quantile as the cut point. For simplicity, we will be assuming an equal number of categorical levels for all categorical variables (set equal to 4). Each data set was generated with 10 variables and five factors were systematically manipulated.

The first factor, the number of clusters in the data sets, was examined at three levels, $K = 2$, 4, and 6. The second factor, the number of observations, was evaluated at three levels, $N = 200$, 500 and 1000, corresponding to a small, moderately large and large sample size in the social and behavioral sciences. The third factor, density of the clusters, was tested at three levels: (a) an equal number of observations in each cluster; (b) 10% of the observations in one cluster and the remaining observations equally divided across the remaining clusters (c) 60% of the observations in one cluster and the remaining observations equally divided across the remaining clusters. The fourth factor, overlap of clusters, assumed values of 0.1%, 1.0%, and 2.0%, corresponding to small, moderate and significant overlap (see Shireman, Steinley & Brusco 2016, for a discussion of overlap in real data sets compared to simulated data sets). The fifth factor was proportions of different variable types in the data and had three conditions (a) 70% categorical, (b) an equal number of continuous and categorical, (c) 70% continuous variables. This resulted in a $3 \times 3 \times 3 \times 3 \times 3 = 243$ distinct data scenarios. In addition, ten replications were made of each scenario, resulting in 2,430 data sets. The ability of each procedure to return the true cluster structure was measured by ARI.

Table 3. Agreement between methods (ARI values) and mean cluster recovery in the analysis of simulated data sets

| **Method** | **(1)** | **(2)** | **(3)** | **(4)** | **(5)** | **Mean ARI** |
|---|---|---|---|---|---|---|
| **(1) Gower/PAM** | - | | | | | .484 |
| **(2) K-prototypes** | .40 | - | | | | .518 |
| **(3) Mixed K-means** | .47 | .49 | - | | | .453 |
| **(4) Mixed hierarchical T** | .36 | .45 | .33 | - | | .517 |
| **(5) Mixed hierarchical B** | .42 | .45 | .27 | .54 | - | .579 |

*Note:* All correlations are significant at $p < .0001$.

### 4.2.2 Overall performance of methods

The correlations between the cluster recovery of the methods and mean cluster recovery, in terms of ARI, are presented in Table 3. None of the correlations are large enough (the largest being Cor(mixed hierarchical B, mixed hierarchical T) = .54 to allow the assertion that one method may serve as a substitution or predictor for another method (for instance, both mixed hierarchical B, mixed hierarchical T only account for 29.2% of the variance in the other). The best performing method was mixed hierarchical B, followed by K-prototypes and mixed hierarchical T, whereas Gower/PAM and mixed K-means were the worst performing ones.

After examining the overall performance of the methods, it is informative to determine if method performance is dependent upon specific situations (i.e., performance varies with the factor levels). The individual performances are examined by the levels of each factor. A repeated-measures ANOVA (see Table 4) was conducted on the true cluster structure recovery. The between data sets effects can be thought of as the influence of the design factors across all clustering methods. To simplify the discussion, only main effects are modeled and discussed. Furthermore, given the large sample size, it was expected that most factors would be statistically significant; therefore, all effects were evaluated with respect to their estimated effect sizes. Effect size was evaluated using partial eta squared (partial $\eta^2$), calculated as the division of the sum of squares of the effects ($SS_{Effect}$) by both the $SS_{Effect}$ and the sum of squares of the error ($SS_{Error}$).

Cluster overlap had the largest effect on cluster recovery. Overall, as the overlap of clusters increased from 0.1% to 2%, the average recovery decreased, going from an ARI = .67 to an ARI = .36. As the number of clusters varied from 4, 6, and 8 while the % of categorical variables varied from 70%, 50%, 30%, there was very little effect on cluster recovery, with ARI values of .57, .50, .46 and .56, .46, 51, respectively. Cluster density had also a small effect on cluster recovery with ARI values equal to .55, .47 and .51 for equally-sized clusters, 10% and 60% of the observations in a single cluster, respectively. Overall, sample size (200, 500 or 1000 observations) did not have a significant effect on cluster recovery (.52, .50 and .50, respectively).

Table 4. Repeated measures ANOVA for five clustering methods on ARI (factors are ordered by decreasing effect size, $\eta^2$).

| Effect | Source | df | SS | *F* | partial $\eta^2$ |
|---|---|---|---|---|---|
| Between data | | | | | |
| sets effects | overlap | 2 | 183.23 | 1603.3 | .143 |
| | number of clusters | 2 | 22.79 | 199.4 | .018 |
| | categorical vars % | 2 | 21.22 | 185.7 | .017 |
| | density | 2 | 12.26 | 107.3 | .010 |
| | sample size | 2 | 1.15 | 10.1 | .001 |
| | Error | 2419 | 138.23 | | |
| Within data sets | Method (M) | 1 | 21.29 | 196.8 | .017 |
| effects | M*categorical vars | 2 | 7.09 | 47.46 | .006 |
| (univariate tests) | M*overlap | 2 | 1.45 | 14.41 | .001 |
| | M*number of clusters | 2 | 1.06 | 9.8 | .001 |
| | M*sample size | 2 | 0.55 | 1.27 | .000 |
| | M*density | 2 | 0.48 | 3.66 | .000 |
| | Error | 105128 | 318.16 | | |

Based on the lower half of Table 4 (within data sets effects) we determine which methods are effective under which conditions. Although there is a significant difference between the five methods in terms of cluster recovery, the effect size is small to moderate. Table 5 shows the ARI values of the five methods by all factors. The percentage of categorical variables in the data, cluster overlap, number of clusters, sample size and cluster density appear to have similar effects on all methods. In other words, Mixed hierarchical B is the best performing method in all situations.

Table 5. Cluster recovery of five clustering methods by cluster overlap, number of clusters, percentage of categorical variables, cluster density and sample size (values of the Adjusted Rand Index)

| Factor | Level | Gower/PAM | K-prototypes | Mixed K-means | Mixed Hierarch T | Mixed Hierarch B |
|---|---|---|---|---|---|---|
| Overlap (%) | 0.1 | .646 | .671 | .592 | .691 | .758∗ |
| | 1 | .461 | .497 | .432 | .486 | .549∗ |
| | 2 | .344 | .344 | .336 | .373 | .429∗ |
| Number of clusters | 2 | .544 | .566 | .494 | .585 | .647∗ |
| | 4 | .473 | .507 | .445 | .504 | .574∗ |
| | 6 | .434 | .482 | .420 | .462 | .514∗ |
| Categorical vars (%) | 70 | .499 | .598 | .529 | .569 | .609∗ |
| | 50 | .450 | .427 | .369 | .491 | .555∗ |
| | 30 | .501 | .529 | .461 | .491 | .572∗ |
| Density (%) | 10 | .442 | .466 | .411 | .487 | .546∗ |
| | 60 | .487 | .529 | .458 | .508 | .572∗ |
| | Equal | .522 | .559 | .490 | .555 | .617∗ |
| Sample size | 200 | .466 | .507 | .441 | .507 | .577∗ |
| | 500 | .485 | .509 | .447 | .510 | .577∗ |
| | 1000 | .499 | .539 | .451 | .532 | .582∗ |

∗Best performing method within each level of all factors.

## 5 Conclusion

In order to decide about appropriate cluster analysis methodology, researchers need to think about what data analytic characteristics the clusters they are aiming at are supposed to have. Hierarchical clustering approaches have been studied and used for decades, are easy to implement and result in an attractive tree-based representation. These advantages come at the cost of lower efficiency, as they do not scale well with a time complexity of at least $O(n^2)$. Most hierarchical clustering algorithms can only handle data that contain either numeric or categorical variables. This work, presented a direct extension of hierarchical clustering to handle nominal, ordinal and interval variables.

Ward's hierarchical clustering based on barycentric coding allows to transform continuous variables, as well as ordinal variables into fuzzy-coded categorical variables. In contrast with the typical discretization of continuous variables, that can lead to significant information loss, barycentric coding is exact and sensitive, even at the smallest differences between the values of a continuous variable. It can also maintain ordinal information in Likert-type scales. This information is lost when alternative fuzzy coding schemes are used, where ordinal variables are always treated as nominal (e.g., Aşan and Greenacre 2011). The transformed data matrix can be considered as a kind of contingency table or a generalization of the dummy (0-1) recoding of categorical variables, making it compatible with Ward's hierarchical clustering using the chi-square distance. This distance has appealing properties and is widely used in multivariate analysis. Columns (variables) with larger weights contribute more to the overall similarity between observations, which is often desirable in practice.

The specific characteristics of the proposed approach have been investigated on both real and simulated data sets. Not surprisingly, results indicated a higher degree of agreement with hierarchical clustering based on fuzzy coding than with other methods. Results on simulated data further revealed that hierarchical clustering based on barycentric coding performed better than alternative methods across all factors examined. However, its performance deteriorated significantly with increasing cluster overlap and number of clusters, factors that also affected all the alternative methods under comparison. On the contrary, the method performed almost equally well on small, moderately large and large sample sizes, although the increase of the data set size in terms of number of columns as a result of the recoding step, was expected to have a negative impact when sample size was small. An obvious limitation stems from the simulation study; mixed-type data were generated after discretization of continuous variables. Constructing an artificial mixed-type data set with a cluster structure that depends on both continuous and categorical variables is still a strenuous task, at least for more than two clusters and specific degree of cluster overlap (see Foss & Markatou 2018, for a discussion). Another limitation is that the effect of the number of fuzzy-coded variables on the proposed method's performance was not assessed via the simulation study. This remains also an open issue when fuzzy coding with any membership function is used to recode continuous data into ordered categories. All in all, the literature on clustering mixed-type data is still relatively sparse and would benefit greatly from further methodological developments and benchmarking studies.

**References**

Ahmad, A. & Dey, L. (2007). A k-mean clustering algorithm for mixed numeric and categorical data, *Data Knowl. Eng.*, *63*(2), 503–527.

Ahmad, A., & Khan, S. S. (2019). Survey of State-of-the-Art Mixed Data Clustering Algorithms. *IEEE Access*, *7*, 31883-31902.

Aşan, Z., & Greenacre, M. (2011). Biplots of fuzzy coded data. *Fuzzy sets and Systems, 183*(1), 57-71.

Benzécri J.-P. (1973). *L'Analyse des Données*, Tome 1: La Taxinomie, Dunod, Paris.

Brudvig, S., Brusco, M. J., & Cradit, J. D. (2019). Joint selection of variables and clusters: recovering the underlying structure of marketing data. *Journal of Marketing Analytics, 7*(1), 1-12.

Escofier, B. (1979). Traitement simultané de variables qualitatives et quantitatives en analyse factorielle. *Cahiers de l'Analyse des Données, 4*(2), 137–146.

Everitt, B. S. (1993). *Cluster analysis*. 3rd ed. N. York: Halsted Press.

Campbell, N. A., & Mahon, R. J. (1974). A multivariate study of variation in two species of rock crab of the genus Leptograpsus. *Australian Journal of Zoology*, *22*(3), 417–425.

Chae, S.-S., Kim, J.-M., Yang, W.-Y. (2006). Cluster analysis with balancing weight on mixed-type data, Communications for Statistical Applications and Methods, *13*(3), 719–732.

Chu, S. (2001). Pricing the C's of Diamond Stones. *Journal of Statistics Education*, *9*(2).

Foss, A. H., & Markatou, M. (2018). kamila: Clustering mixed-type data in R and Hadoop. *Journal of Statistical Software, 83*(1), 1–44.

Greenacre, M., & Hastie, T. (1987). The geometric interpretation of correspondence analysis. *Journal of the American Statistical Association*, *82*(398), 437-447.

Greenacre, M. (2013). Fuzzy coding in constrained ordinations. *Ecology*, 94, 280–286.

Greenacre, M. (2017). *Correspondence Analysis in Practice.* Third edition. Chapman and Hall/CRC.

Gower, J. C. (1971). A general coefficient of similarity and some of its properties. *Biometrics*, 857–871.

Hennig, C., & Liao, T. F. (2013). How to find an appropriate clustering for mixed-type variables with application to socio-economic stratification. *Journal of the Royal Statistical Society: Series C (Applied Statistics)*, *62*(3), 309–369.

Hsu, C.-C., Chen, C.-L., Su, Y-W (2007). Hierarchical clustering of mixed data based on distance hierarchy. *Information Sciences*, *177*, 4474–4492.

Huang, Z. (1998). Extensions to the k-means algorithm for clustering large data sets with categorical values. *Data mining and knowledge discovery, 2*(3), 283–304.

Hubert, L., & Arabie, P. (1985). Comparing partitions. *Journal of Classification, 2*(1), 193–218.

Hunt, L., & Jorgensen, M. (2011). Clustering mixed data. *WIREs Data Mining and Knowledge Discovery, 1*(4), 352–361. https://doi.org/10.1002/widm.33.

Lê, S., Josse, J., Husson, F. (2008). FactoMineR: An R package for multivariate analysis. *Journal of Statistical Software*, 25, 1-18.

Le Roux, B., & Rouanet, H. (2004). *Geometric data analysis: from correspondence analysis to structured data analysis*. Springer Science & Business Media.

Lebart, L., Morineau, A., Warwick, K. M. (1984). *Multivariate descriptive statistical analysis; correspondence analysis and related techniques for large matrices*. New York: John Wiley and Sons.

Li, C., & Biswas, G. (2002). Unsupervised learning with mixed numeric and nominal data. *IEEE Transactions on Knowledge & Data Engineering*, 4, 673-690.

Kaufman, L. & Rousseeuw, P. J. (1990). *Finding groups in data: an introduction to cluster analysis.* Vol. 344. John Wiley & Sons.

Maechler, M., Rousseeuw, P., Struyf, A., Hubert, M., Hornik, K. (2018). cluster: Cluster analysis basics and extensions [Computer Software manual]. R package version 2.0.7-1. Retrieved from https://CRAN.R-project.org/package=cluster.

Melnykov, V., & Maitra, R. (2010). Finite mixture models and model-based clustering. *Statistics Surveys*, *4*, 80–116.

McParland, D., Phillips, C. M., Brennan, L., Roche, H. M., Gormley, I. C. (2017). Clustering high-dimensional mixed data to uncover subphenotypes: Joint analysis of phenotypic and genotypic data, *Statistics in Medicine*, *36*(28), 4548–4569.

Melnykov, V., Chen, W.-C., & Maitra, R. (2012). MixSim: An R package for simulating data to study performance of clustering algorithms. *Journal of Statistical Software*, *51*(12), 1–25.

Morlini, I. I., & Zani, S. (2010). Comparing approaches for clustering mixed mode data: An application in marketing research. In *Data Analysis and Classification*. Berlin, Germany: Springer, 2010, pp. 49–57.

Moschidis, O., & Chadjipadelis, T. (2017). A Method for Transforming Ordinal Variables. In *Data Science* (pp. 285–294). Springer, Cham.

Moschidis, O. (2015). Unified coding of qualitative and quantitative variables and their analysis with ascendant hierarchical classification, *International Journal of Data Analysis Techniques and Strategies, 7*(2), 114–128.

Murtagh, F., & Legendre, P. (2014). Ward's Hierarchical Agglomerative Clustering Method: Which Algorithms Implement Ward's Criterion? *Journal of Classification, 31*(3), 274–295.

Pathberiya, H. A. (2016). DisimForMixed: Calculate dissimilarity matrix for dataset with mixed attributes [Computer software manual]. R package version 0.2. Retrieved from https://CRAN.R-project.org/package=DisimForMixed.

Shireman, E. M., Steinley, D., & Brusco, M. J. (2016). Local optima in mixture modeling. *Multivariate Behavioral Research*, *51*(4), 466–481.

Szepannek, G. (2017). clustMixType: k-prototypes clustering for mixed variable-type data [Computer software manual]. R package version 0.1-29. Retrieved from https://CRAN.R-project.org/package=clustMixType.

van de Velden, M., Iodice D'Enza, A., Markos, A. (2019). Distance-based clustering of mixed data. *WIREs Computational Statistics* (Advanced Review, online first) https://doi.org/10.1002/wics.1456.

Van Rijckevorsel, J. L. A. (1988). Fuzzy coding and B-splines. Component and correspondence analysis. *Dimension reduction by functional approximation*, 33-54.

Vichi, M., Vicari, D., Kiers, H. A. L. (2019). Clustering and dimension reduction for mixed variables. *Behaviormetrika.* doi:10.1007/s41237-018-0068-6.

## Appendix – R code for barycentric coding

```
barycentric <- function(A, cats = 5, con = TRUE)
{
#Coding of a continuous variable into categorical via barycentric recoding
#Example
#A = c(40, 33, 32.5, 32, 55.2, 60.1, 32)
#barycentric(A, 3, con = TRUE)
#converts a continuous variable A into a fuzzy-coded categorical with 3 categories
  #use con = TRUE for continuous
  #use con = FALSE for Likert-type ordinal
  if (con == TRUE) {
    minA = min(A)
    maxA = max(A)
    absdist = dist(A,"manhattan")
    dmin = min(apply(as.matrix(absdist), 1, FUN = function(x) {min(x[x > 0])}))
    all = round((maxA - minA) / dmin + 1)
    Acat={}
    for (i in 1:length(A)) {
      Acat[i] = trunc((A[i] - minA) / dmin) + 1
    }
    A = Acat
  }
  #Apply barycentric coding
  maxA = max(A)
  Ni = length(A)
```

```
  pLOW = 1/2
  k = 0
  inVals = cats

  R = matrix(0,Ni,inVals)
  p00 = rep(0,inVals+1)
  #for (j in 0:(Nj-1)) {
  pHIGH = maxA + 1/2
  p00[1] = pLOW
  for (cnt in 1:cats) {
    p00[cnt+1] = p00[cnt] + (pHIGH - pLOW) / inVals
  }
  for (i in 1:Ni) {
    eee = Gen_split_w_center(A[i],p00)
    for (cnt in 1:(inVals)) {
      R[i,cnt] = eee[cnt]
    }
  }
  #}
  return(R)

}

Low_level_split <- function(s, sx, a, b) {
  Low_level_split = sx * (b - s) / (b - a)
}
Gen_split_w_center <- function(s, p) {
  x= rep(0,length(p)-1)
  flag = 0
  for (u in 1:(length(p)-1)) {
```

```
  if (s == p[u]) {
    flag = 1
    Base = u
    base_a = Base - 1
  } else if ((s > p[u]) & (s < p[u+1])) {
    Base = u
    base_a = Base
  }
}
a = p[base_a]
b = p[Base + 1]
xa = Low_level_split(s, 1, a, b)
xb = 1 - xa
if (flag == 1)
  a = p[Base]
if ((Base +1) <= (length(p)-1)) {
  for (i in (Base + 1):(length(p)-1)) {
    a2 = (b + a) / 2
    b2 = p[i + 1]
    xa2 = Low_level_split(b, xb, a2, b2)
    xb2 = xb - xa2
    x[i-1] = xa2
    a = b
    b = b2
    xb = xb2
  }
}
x[length(p)-1] = xb
a = p[base_a]
b = p[Base + 1]
```

```
  xa = Low_level_split(s, 1, a, b)
  xb = 1 - xa
  if (flag == 1)
    b = p[Base]
  if (base_a >= 2) {
    for (k in seq(base_a, 2, by=-1))
    {
      b2 = (a + b) / 2
      a2 = p[k-1]
      xa2 = Low_level_split(a, xa, a2, b2)
      xb2 = xa - xa2
      x[k] = x[k] + xb2
      b = a
      a = a2
      xa = xa2
    }
  }
  x[1] = x[1] + xa

  return(x)
}
```